\documentclass[11pt,letterpaper]{article}

\IfFileExists{latexml.sty}{\usepackage{latexml}}{\newif\iflatexml\latexmlfalse}

\usepackage[T1]{fontenc}
\usepackage[utf8]{inputenc}
\iflatexml
\newcommand{\crlt}{}\newcommand{\hlv}{}
\else
\usepackage{lmodern}
\usepackage[default,lining]{lato}
\DeclareFontFamily{T1}{LatoW}{}
\DeclareFontShape{T1}{LatoW}{m}{n}{<-> Lato-Regular-T1-TLF--base}{}
\DeclareFontShape{T1}{LatoW}{m}{it}{<-> Lato-Italic-T1-TLF--base}{}
\DeclareFontShape{T1}{LatoW}{b}{n}{<-> Lato-Bold-T1-TLF--base}{}
\DeclareFontShape{T1}{LatoW}{b}{it}{<-> Lato-BoldItalic-T1-TLF--base}{}
\DeclareFontShape{T1}{LatoW}{m}{sl}{<-> ssub * LatoW/m/it}{}
\DeclareFontShape{T1}{LatoW}{b}{sl}{<-> ssub * LatoW/b/it}{}
\DeclareFontShape{T1}{LatoW}{bx}{n}{<-> ssub * LatoW/b/n}{}
\DeclareFontShape{T1}{LatoW}{bx}{it}{<-> ssub * LatoW/b/it}{}
\DeclareFontShape{T1}{LatoW}{bx}{sl}{<-> ssub * LatoW/b/it}{}
\DeclareFontFamily{TS1}{LatoW}{}
\DeclareFontShape{TS1}{LatoW}{m}{n}{<-> ssub * lato-TLF/m/n}{}
\DeclareFontShape{TS1}{LatoW}{m}{it}{<-> ssub * lato-TLF/m/it}{}
\DeclareFontShape{TS1}{LatoW}{b}{n}{<-> ssub * lato-TLF/b/n}{}
\DeclareFontShape{TS1}{LatoW}{b}{it}{<-> ssub * lato-TLF/b/it}{}
\DeclareFontShape{TS1}{LatoW}{bx}{n}{<-> ssub * lato-TLF/b/n}{}
\DeclareFontShape{TS1}{LatoW}{bx}{it}{<-> ssub * lato-TLF/b/it}{}

\DeclareFontFamily{T1}{CrltW}{}
\DeclareFontShape{T1}{CrltW}{m}{n}{<-> Crlt-tlf-t1--base}{}
\DeclareFontShape{T1}{CrltW}{m}{it}{<-> Crlt-Italic-tlf-t1--base}{}
\DeclareFontShape{T1}{CrltW}{b}{n}{<-> Crlt-Bold-tlf-t1--base}{}
\DeclareFontShape{T1}{CrltW}{b}{it}{<-> Crlt-BoldItalic-tlf-t1--base}{}
\DeclareFontShape{T1}{CrltW}{bx}{n}{<-> ssub * CrltW/b/n}{}
\DeclareFontShape{T1}{CrltW}{bx}{it}{<-> ssub * CrltW/b/it}{}
\DeclareFontFamily{T1}{CldaW}{}
\DeclareFontShape{T1}{CldaW}{m}{n}{<-> Caladea-Regular-tlf-t1--base}{}
\DeclareFontShape{T1}{CldaW}{b}{n}{<-> Caladea-Bold-tlf-t1--base}{}
\DeclareFontShape{T1}{CldaW}{bx}{n}{<-> ssub * CldaW/b/n}{}
\newcommand{\crlt}{\fontfamily{CrltW}\selectfont}

\newcommand{\hlv}{\fontfamily{phv}\selectfont}
\usepackage{textcomp}
\usepackage[protrusion=false,expansion=false]{microtype}
\DisableLigatures[f]{encoding = *, family = *}
\input{glyphtounicode}\pdfgentounicode=1
\pdfglyphtounicode{fi}{0066 0069}\pdfglyphtounicode{f_i}{0066 0069}\pdfglyphtounicode{fl}{0066 006C}\pdfglyphtounicode{f_l}{0066 006C}
\pdfglyphtounicode{ff}{0066 0066}\pdfglyphtounicode{f_f}{0066 0066}\pdfglyphtounicode{ffi}{0066 0066 0069}\pdfglyphtounicode{f_f_i}{0066 0066 0069}
\pdfglyphtounicode{uniFB00}{0066 0066}\pdfglyphtounicode{uniFB01}{0066 0069}\pdfglyphtounicode{uniFB02}{0066 006C}\pdfglyphtounicode{uniFB03}{0066 0066 0069}\pdfglyphtounicode{uniFB04}{0066 0066 006C}
\fi

\usepackage[letterpaper,left=56.16bp,textwidth=499.68bp,top=79.2bp,textheight=658.8bp,
            headheight=0bp,headsep=0bp,footskip=0bp]{geometry}
\usepackage[table]{xcolor}
\usepackage{graphicx}
\usepackage{array}
\usepackage{longtable}
\usepackage{titlesec}
\iflatexml\else\usepackage[titles]{tocloft}\fi
\iflatexml\else\usepackage{eso-pic}\fi
\usepackage[document]{ragged2e}
\iflatexml\else\usepackage[most]{tcolorbox}\fi
\usepackage[compress,nosort]{cite}
\usepackage{url}
\usepackage{etoolbox}
\usepackage[hidelinks]{hyperref}
\mathchardef\citeprepenalty=0

\hypersetup{
  pdftitle={Distance from the Truth: Why Fragmented Merchant Underwriting Misses Risk, and What Source-Attributable Evidence Changes},
  pdfauthor={Russell Lessard, Jeremy Blackburn, Tim Hines, Matthew Koepp, Jason Patrick},
  pdfsubject={A merchant-underwriting application of the ChainIT Transaction Truth Architecture (CIT-DTT-WP-2.1)},
  pdfkeywords={merchant underwriting, know your business, customer due diligence, synthetic identity fraud, evidence provenance, source lineage, organizational authority, payment facilitators, ChainIT}
}

\definecolor{body}{HTML}{24313A}
\definecolor{dgreen}{HTML}{4F6228}
\definecolor{subgreen}{HTML}{76923C}
\definecolor{cgray}{HTML}{5C6872}
\definecolor{lgray}{HTML}{A6A6A6}
\definecolor{clink}{HTML}{126B78}
\definecolor{author}{HTML}{181E2A}
\definecolor{hdr}{HTML}{9BBB59}
\definecolor{rowalt}{HTML}{EAF1DD}
\definecolor{trule}{HTML}{C2D69B}
\definecolor{boxfill}{HTML}{EDF4F5}

\renewcommand{\normalsize}{\fontsize{11bp}{14.24bp}\selectfont}
\newcommand{\tabsize}{\fontsize{10bp}{12.54bp}\selectfont}
\AtBeginDocument{\normalsize\color{body}}
\newcommand{\gbf}[1]{\textbf{\color{dgreen}#1}}

\newif\ifboxlast\newif\iftablelast\newif\ifqhead
\iflatexml\else\AddToHook{para/begin}{\global\boxlastfalse\global\tablelastfalse\global\qheadfalse}\fi
\titleformat{\section}{\raggedright\crlt\bfseries\fontsize{18bp}{21.6bp}\selectfont\color{dgreen}}{\rule{0bp}{17.04bp}\thesection.\ }{0bp}{\rule{0bp}{17.04bp}}
\makeatletter\titleformat{\subsection}{\if@nobreak\vspace{0.8bp}\fi\raggedright\crlt\bfseries\fontsize{12.5bp}{14.98bp}\selectfont\color{subgreen}}{\thesubsection\ }{0bp}{}\makeatother
\titlespacing*{\section}{0bp}{\ifboxlast 4.9bp\else\iftablelast 9bp\else 8.5bp\fi\fi}{4.3bp}
\makeatletter\titlespacing*{\subsection}{0bp}{\ifqhead 1.95bp\else\iftablelast 3.5bp\else 5.05bp\fi\fi}{-0.5bp}\makeatother
\newcommand{\qhead}[1]{\global\qheadtrue\subsection*{#1}}
\newcommand{\unnumbered}[1]{\section*{#1}\addcontentsline{toc}{section}{\protect\toccaps{#1}}}
\iflatexml
\newcommand{\toccaps}[1]{#1}
\else
\DeclareRobustCommand{\toccaps}[1]{\texorpdfstring{\MakeUppercase{#1}}{#1}}
\fi

\iflatexml\else

\newsavebox{\tocnum}
\renewcommand{\cftsecpresnum}{\global\setbox\tocnum\hbox\bgroup}
\renewcommand{\cftsecaftersnum}{.\egroup}

\makeatletter\renewcommand{\@tocrmarg}{14bp}\renewcommand{\@pnumwidth}{1.5em}\makeatother
\fi

\iflatexml
\newcommand{\dttbox}[1]{\begin{quote}\textbf{#1}\end{quote}}
\else
\newcommand{\dttbox}[1]{%
  \par\kern-\prevdepth
  \begin{tcolorbox}[enhanced,frame hidden,interior hidden,boxrule=0bp,arc=0bp,boxsep=0bp,colupper=dgreen,
    left skip=6bp,right skip=6bp,left=1.5bp,right=1.5bp,top=0bp,bottom=0bp,
    before skip=10.3bp,after skip=5.2bp,
    underlay={\fill[boxfill](frame.south west) rectangle (frame.north east);},
    before upper={\RaggedRight\fontsize{11.5bp}{16.08bp}\selectfont\rule{0bp}{11.28bp}}]
  #1\rule[-4.56bp]{0bp}{0bp}%
  \end{tcolorbox}\prevdepth=0bp\global\boxlasttrue}
\fi

\arrayrulecolor{trule}
\newcommand{\cellpadtop}{\rule{0bp}{11.9bp}}
\newcommand{\cellpadbot}{\rule[-4.55bp]{0bp}{0bp}}
\newcolumntype{P}[1]{>{\RaggedRight\arraybackslash\tabsize\cellpadtop}p{#1}<{\cellpadbot}}
\newcommand{\hc}[1]{\color{white}#1}
\newcommand{\hdrrow}{\rowcolor{hdr}}
\newcommand{\conthead}{\noalign{{\color{trule}\hrule height\arrayrulewidth}\kern\dimexpr\topskip-\arrayrulewidth\relax\setbox0\hbox{}\dp0=\dimexpr\arrayrulewidth-\topskip\relax\box0}}
\newcommand{\hdrend}{\arrayrulecolor{trule}}
\iflatexml
\newcommand{\dtabstart}{}
\else
\newcommand{\dtabstart}{\par\penalty0\kern-\prevdepth\kern9.9bp\rowcolors{2}{white}{rowalt}\arrayrulecolor{hdr}}
\AfterEndEnvironment{longtable}{\prevdepth=0bp\vskip-3.9bp\global\tablelasttrue}
\fi

\iflatexml
\newcommand{\refurl}[1]{\url{#1}}
\newcommand{\refurlb}[1]{\url{#1}}
\newcommand{\refurlsplit}[3]{\url{#1}}
\newcommand{\refurlsmall}[1]{\url{#1}}
\else
\makeatletter
\newcommand{\refurlfont}{\fontsize{9.5bp}{11.04bp}\selectfont\color{clink}}
\newcommand{\refurl}[1]{\vadjust pre{\kern0.6bp}{\refurlfont\href{#1}{\refword{#1}}}}
\newcommand{\refurlsplit}[3]{\vadjust pre{\kern0.6bp}{\refurlfont\href{#1}{\refword{#2}}}\break\vadjust pre{\kern0.48bp}{\refurlfont\href{#1}{\refword{#3}}}}
\newcommand{\refurlb}[1]{\vadjust pre{\kern0.6bp}{\refurlfont\href{#1}{\refseg@#1-\@nil}}}
\newcommand{\refurlsmall}[1]{{\color{clink}\href{#1}{\nolinkurl{#1}}}}
\def\refseg@#1-#2\@nil{\ifx\relax#2\relax\ifx\relax#1\relax\else\refword{#1}\fi\else\refword{#1-}\penalty-1 \refseg@#2\@nil\fi}
\newcommand{\refword}[1]{\setbox0\hbox{\nolinkurl{#1}}\rlap{\rule[-2.16bp]{\wd0}{0.48bp}}\box0}
\makeatother
\fi
\iflatexml\else
\makeatletter
\renewenvironment{thebibliography}[1]
  {\list{\@biblabel{\@arabic\c@enumiv}}{\leftmargin0bp\labelwidth0bp\labelsep0bp\itemindent0bp\listparindent0bp
     \itemsep7.2bp\parsep0bp\topsep0bp\partopsep0bp\usecounter{enumiv}}\RaggedRight}
  {\endlist}
\makeatother
\fi

\iflatexml\else
\AddToShipoutPictureBG{\setlength{\unitlength}{1bp}%
  \ifnum\value{page}=1
    \put(0.8498,-25.0499){\includegraphics[width=759.6872bp,height=808.35bp]{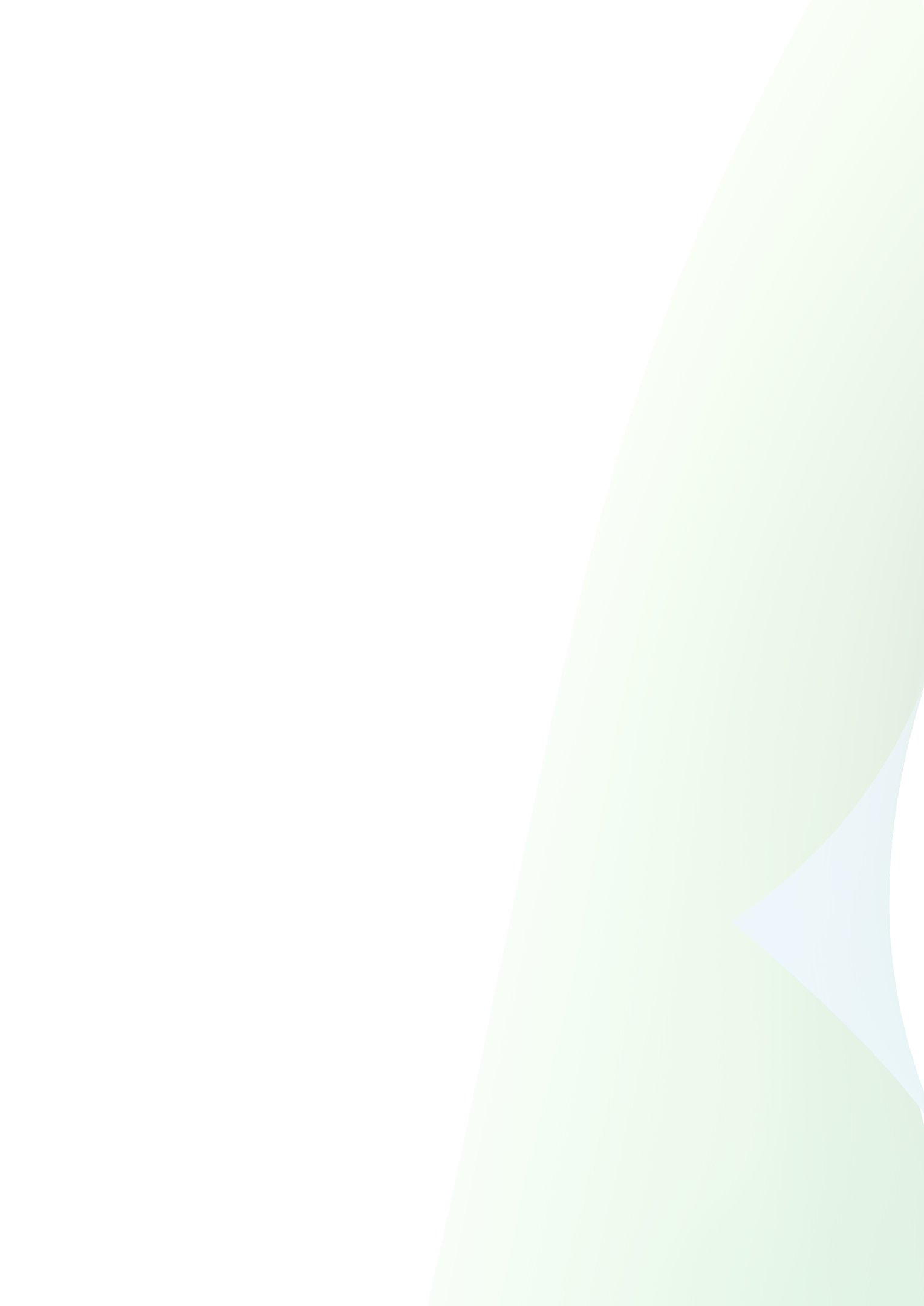}}%
    \put(414.75,737.1499){\includegraphics[width=134.0002bp,height=38bp]{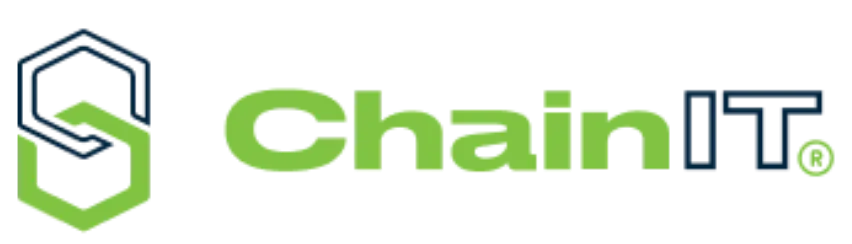}}%
  \else
    \put(-6.5,-7.85){\includegraphics[width=679.6883bp,height=790.35bp]{figures/page-background.png}}%
    \put(425,735.5){\includegraphics[width=134.0002bp,height=38bp]{figures/chainit-logo.png}}%
    \put(56.16,24){\fontsize{8bp}{9.5bp}\selectfont\color{cgray}CIT-DTT-WP-2.1 \textbar{} 29 September 2026}%
    \put(291.57,24){\makebox[0bp]{\fontsize{11bp}{13bp}\selectfont\color{body}-{\fontsize{8bp}{9.5bp}\selectfont\color{cgray}\ }\thepage\ - }}%
    \put(431.52,24.48){\fontsize{8bp}{9.5bp}\selectfont\color{lgray}CHAINIT\ \ \textbar\ \ Distance from the Truth}%
  \fi}
\fi

\title{Distance from the Truth: Why Fragmented Merchant Underwriting Misses Risk, and What Source-Attributable Evidence Changes}
\newcommand{\affil}{ChainIT, Utah, United States (\url{https://chainit.com})}%
\author{Russell Lessard\thanks{\affil} \and Jeremy Blackburn\thanks{\affil} \and Tim Hines\thanks{\affil} \and Matthew Koepp\thanks{\affil} \and Jason Patrick\thanks{\affil}}
\date{29 September 2026}

\iflatexml\else
\renewenvironment{abstract}{\section*{Abstract}}{\par}
\fi

\begin{document}

\iflatexml
\maketitle
\else
\thispagestyle{empty}
\pagestyle{empty}
\newcommand{\coverc}[3]{\put(306,#1){\makebox[0bp]{#2#3}}}
\newcommand{\coverl}[3]{\put(56.16,#1){\parbox[t]{499.68bp}{\RaggedRight\leavevmode#2#3}}}
\begin{picture}(0,0)\setlength{\unitlength}{1bp}%
\end{picture}%
\AddToShipoutPictureFG*{\setlength{\unitlength}{1bp}%
\coverc{678.96}{\crlt\fontsize{12bp}{14.4bp}\selectfont\color{dgreen}}{\textls[63]{CHAINIT}}%
\coverc{659.28}{\fontsize{9bp}{11bp}\selectfont\color{cgray}}{\textls[-13]{MERCHANT UNDERWRITING /} \url{https://chainit.com}}%
\coverc{515.76}{\bfseries\fontsize{36bp}{43.2bp}\selectfont\color{dgreen}}{Distance from the Truth}%
\coverc{465.84}{\fontsize{19bp}{22.75bp}\selectfont\color{dgreen}}{Why fragmented merchant underwriting misses risk,}%
\coverc{441.36}{\fontsize{19bp}{22.75bp}\selectfont\color{dgreen}}{and what source-attributable evidence changes}%
\coverc{396.00}{\bfseries\fontsize{11.5bp}{13.8bp}\selectfont\color{dgreen}}{Provenance. Correlation. Persistence.}%
\coverc{371.76}{\fontsize{11bp}{13.25bp}\selectfont\color{dgreen}}{A merchant-underwriting application of the}%
\coverc{357.60}{\fontsize{11bp}{13.25bp}\selectfont\color{dgreen}}{ChainIT \href{https://chainit.com/white-papers/}{Transaction Truth} Architecture}%
\coverc{338.64}{\fontsize{9bp}{11bp}\selectfont\color{cgray}}{Version 2.1 \textbar{} 29 September 2026}%
\coverc{326.40}{\fontsize{9bp}{11bp}\selectfont\color{cgray}}{CIT-DTT-WP-2.1}%
\coverc{288.72}{\hlv\fontsize{10bp}{12bp}\selectfont\color{dgreen}}{Authors}%
\coverc{277.68}{\hlv\fontsize{9bp}{10.8bp}\selectfont\color{author}}{\href{https://chainit.com/about-us/}{Russell Lessard}, Chief Compliance Officer}%
\coverc{266.88}{\hlv\fontsize{9bp}{10.8bp}\selectfont\color{author}}{\href{https://chainit.com/about-us/}{Jeremy Blackburn}, Chief Executive Officer}%
\coverc{256.08}{\hlv\fontsize{9bp}{10.8bp}\selectfont\color{author}}{\href{https://chainit.com/about-us/}{Tim Hines}, Chief Product Officer}%
\coverc{245.28}{\hlv\fontsize{9bp}{10.8bp}\selectfont\color{author}}{\href{https://chainit.com/about-us/}{Matthew Koepp}, Chief Technology Officer}%
\coverc{234.48}{\hlv\fontsize{9bp}{10.8bp}\selectfont\color{author}}{\href{https://chainit.com/about-us/}{Jason Patrick}, Chief Commercial Officer}%
\coverl{155.52}{\bfseries\fontsize{11bp}{13.25bp}\selectfont\color{dgreen}}{Trust Nothing. Verify Everything. With ChainIT.}%
\coverl{136.80}{\fontsize{9bp}{12.48bp}\selectfont\color{cgray}}{For merchant-risk leaders, acquirers, payment facilitators, independent sales organizations, sponsor banks, and third-party risk teams.}%
\coverl{106.08}{\fontsize{9bp}{12.48bp}\selectfont\color{cgray}}{This paper describes an evidence and control architecture, not legal advice, a compliance certification, or a guarantee against fraud or loss. Deployment scope, provider access, and operational availability are implementation-specific.}%
\coverl{75.12}{\fontsize{9bp}{12.48bp}\selectfont\color{cgray}}{\textcopyright{} 2026 Scientia Potentia Est II, LLC. All rights reserved. Distributed by arXiv under the arXiv.org perpetual, non-exclusive license.}%
}
\null\clearpage
\pagestyle{empty}
\fi
\color{body}

\begin{abstract}
Merchant underwriting becomes vulnerable when evidence is evaluated without enough visibility into its source, freshness, relationship to the applicant, or meaning for the proposed activity. Individually favorable checks do not necessarily establish that the same business is operating the merchant account, that its representative may bind it, or that its settlement instructions are authorized. We describe three failure modes of fragmented underwriting: distance from the source, the gap between a decision and the evidence needed to justify it; the correlation gap, in which material relationships among the person, legal entity, owners or controllers, representative, business activity, and settlement destination are not evaluated when each signal is considered alone; and point-in-time controls, in which a historical verification is treated as present eligibility. We map these failure modes to identity-proofing, customer-due-diligence, sanctions, third-party-risk, card-network, and Nacha requirements, and propose a merchant-underwriting application of the ChainIT \href{https://chainit.com/white-papers/}{Transaction Truth} Architecture built on provenance, correlation, and persistence. The application keeps a logically unified evidence record that identifies what was checked, the subject and relationship involved, the relevant times, the applicable policy, the decision, and any exceptions, and it keeps evidence validation, applicant verification, organizational authority, and the institution\textquotesingle{}s risk judgment separate. We also set out a proposed evidence and decision receipt, a defined 90-day shadow enrollment pilot with adjudicated outcome measures and deliberate control-failure tests, and a ten-question self-assessment. Proposed controls, illustrative scenarios, and pilot measures are identified as such and are not represented as observed production results.
\end{abstract}


{\fontsize{9bp}{12.48bp}\selectfont\color{cgray}This paper was first released as a ChainIT white paper (CIT-DTT-WP-2.1). The companion papers, \emph{\href{https://chainit.com/insights/chainit-launches-provable-authority-as-u-s-generative-ai-enabled-fraud-losses-are-projected-to-reach-40-billion-by-2027/}{Provable Authority}} (arXiv:\href{https://arxiv.org/abs/2610.22899}{2610.22899}), \emph{\href{https://chainit.com/white-papers/}{Provable Compliance}}~\cite{c1} and \emph{\href{https://chainit.com/white-papers/}{Transaction Truth}} (arXiv:\href{https://arxiv.org/abs/2611.24659}{2611.24659}), and other ChainIT white papers are listed at \url{https://chainit.com/white-papers/}.\par}
\clearpage

\iflatexml\else
\tableofcontents
\vspace{20.14bp}
\fi
{\fontsize{9bp}{10.8bp}\selectfont\color{cgray}\textbf{Reading guide.} Sections 1-6 develop the vendor-neutral risk and regulatory argument. Section 7 maps that argument to ChainIT. Sections 8-10 address procurement, validation, and operational self-assessment. Appendix A is a proposed application profile, not a new protocol.\par
\vspace{5.5bp}
\textbf{Source convention.} Numbered references identify external primary authorities. References C1-C3 identify ChainIT architectural and technical source material. Proposed controls, illustrative scenarios, and pilot measures are identified as such and are not represented as observed production results.\par}

\clearpage
\addtocontents{toc}{\protect\vspace*{2.3bp}}%
\unnumbered{Executive Summary}
Merchant underwriting becomes vulnerable when evidence is evaluated without enough visibility into its source, freshness, relationship to the applicant, or meaning for the proposed activity. Individually favorable checks do not necessarily establish that the same business is operating the merchant account, that its representative may bind it, or that its settlement instructions are authorized.

The answer is not simply more vendors, fewer vendors, or more uploaded documents. A well-integrated multi-provider system can be effective; a single-provider system can preserve the same gaps. The relevant design question is whether the institution can reconstruct a defensible decision from attributable evidence and act appropriately when the evidence changes.

\dtabstart
\begin{longtable}{|@{\hspace{5.04bp}}P{155.52bp}@{\hspace{5.28bp}}|@{\hspace{5.04bp}}P{155.52bp}@{\hspace{5.28bp}}|@{\hspace{5.04bp}}P{156.24bp}@{\hspace{5.28bp}}|}
\hline
\hdrrow\hc{PROVENANCE} & \hc{CORRELATION} & \hc{PERSISTENCE}\\\hline
\hdrend
\endfirsthead
\conthead
\endhead
\textbf{What exact fact was established, by which source, through which access path, and with what limitations?} & Do the person, organization, ownership or control information, authority, business activity, and settlement destination form a coherent relationship? & Is the evidence fit for this reliance event, and what happens when status, authority, activity, or source availability changes?\\\hline
\end{longtable}
These principles require a further distinction: \textbf{identity is not authority, and compliance is not creditworthiness}. An authentic business can experience financial distress. A verified employee can lack authority to change its bank account. A favorable sanctions-screening result does not independently permit every transaction. Each conclusion belongs to a defined purpose, policy, and accountable decision owner.

ChainIT applies this model through verified individual and organizational identities, organizational authority records, modular evidence assertions, policy evaluation, and attributable lifecycle records. Within the ChainIT \href{https://chainit.com/white-papers/}{Transaction Truth} Architecture, the ChainIT Compliance Protocol and ChainIT \href{https://chainit.com/insights/chainit-launches-provable-authority-as-u-s-generative-ai-enabled-fraud-losses-are-projected-to-reach-40-billion-by-2027/}{Authority Protocol} remain separate, adjacent protocols. The underwriting application proposed here connects them to the institution\textquotesingle{}s merchant-risk and operational controls; it does not replace the institution\textquotesingle{}s risk appetite or legal responsibilities. \cite{c1}

The practical objective is a \gbf{logically unified evidence record}, not unrestricted central storage of every source document. Such a record identifies what was checked, the subject and relationship involved, the relevant times, the applicable policy, the decision, and any exceptions. Data is retained or referenced according to access rights, security requirements, and the institution\textquotesingle{}s retention schedule.

The recommended adoption path is a defined 90-day shadow enrollment pilot, followed by the observation periods needed for outcome measures. Evaluate evidence completeness, legitimate-merchant friction, adjudicated disagreements, control failures, and fully loaded cost. Do not call a disagreement fraud, or an unprocessed application a prevented financial loss.

\dttbox{\gbf{The standard is not seven green lights. It is an explainable decision about the correct subject, relationship, authority, activity, and moment of reliance.}}
\clearpage
\section[\protect\toccaps{The Problem Acquirers Are Living With}]{The Problem Acquirers Are Living With}
\subsection{The failure is often between otherwise valid facts}
Merchant due diligence combines identification, ownership or control information, business-model assessment, financial analysis, sanctions and other risk controls, contractual onboarding, and monitoring. Card-network standards address merchant screening and continuing oversight; bank supervision also expects attention to the risks of third-party activities. Neither framework makes vendor count a substitute for an effective control process. \cite{r12,r13,r14}

The following threat scenarios explain why isolated checks can be insufficient. They are illustrative control problems, not claims about their relative frequency in any merchant portfolio.

\dtabstart
\begin{longtable}{|@{\hspace{5.04bp}}P{133.20bp}@{\hspace{5.28bp}}|@{\hspace{5.04bp}}P{345.12bp}@{\hspace{5.04bp}}|}
\hline
\hdrrow\hc{Illustrative threat} & \hc{What an isolated check may miss}\\\hline
\hdrend
\endfirsthead
\conthead
\endhead
\textbf{Hijacked legitimate business} & A real registration and tax identifier may be presented by someone who lacks a verified relationship or authority.\\\hline
\textbf{Synthetic or misrepresented business} & Individually plausible attributes may describe no coherent operating business, or describe a different activity from the one proposed.\\\hline
\textbf{Undisclosed processing or transaction laundering} & The entity that passed onboarding may not be the actual seller or activity generating the transactions.\\\hline
\textbf{Concealed ownership or control} & The applicant may omit a relevant relationship; confirming the identity of the disclosed owner does not prove that the disclosure was complete.\\\hline
\textbf{Post-onboarding takeover or payout diversion} & An accepted identity or login may be reused to change settlement instructions without the required current authority.\\\hline
\textbf{Financial or fulfillment failure} & Even a genuine, authorized merchant may be unable to deliver prepaid goods, honor refunds, or absorb obligations.\\\hline
\end{longtable}
\subsection{Use evidence about the problem without overstating it}
FinCEN\textquotesingle{}s January 2024 analysis identified approximately 1.6 million identity-related BSA reports, involving about \$212 billion in suspicious activity, in its analysis of 2021 reporting. Those figures describe reported suspicious activity. They are not confirmed losses, are not specific to merchant onboarding, and do not establish a current merchant-fraud growth rate. They support the importance of identity exploitation as a financial-system concern, not a merchant-market forecast. \cite{r1}

The Federal Reserve\textquotesingle{}s industry definition of synthetic identity fraud is \textquotedbl{}the use of a combination of personally identifiable information (PII) to fabricate a person or entity in order to commit a dishonest act for personal or financial gain,\textquotedbl{} and its 2019 analysis of synthetic identity fraud in the payment system states that \textquotedbl{}[f]raudsters also use synthetic identities to create fake businesses and sign up with merchant processors to obtain credit card terminals and run up charges on fraudulent cards.\textquotedbl{} The case it used as illustration involved more than 7,000 synthetic identities, more than 25,000 credit cards, and, according to the FBI, \textquotedbl{}dozens of sham companies\textquotedbl{} that established merchant accounts and, when those accounts were closed, \textquotedbl{}would apply for new terminals and create new companies.\textquotedbl{} Confirmed losses exceeded \$200 million. The case is dated and the definition is broad; they are cited because they document, from official sources, that merchant onboarding is a target of this fraud type, not as evidence of its current rate. \cite{r25,r26}

Consumer-side figures give the surrounding scale without measuring merchant onboarding itself. The Federal Trade Commission reported three million consumer fraud reports and \$15.9 billion in reported losses for 2025, with more than one million reports involving impersonation; the merchant accounts through which such proceeds are collected must be opened somewhere. TransUnion estimated U.S. lender exposure to synthetic identities on consumer credit products at \$3.2 billion at mid-2024. Both are cited for context: the first is unverified consumer reporting, and the second measures outstanding balances on consumer accounts a vendor\textquotesingle{}s model flags as synthetic, not merchant losses. \cite{r27,r28}

\subsection{Verification and underwriting solve different problems}
The proposed evidence architecture supports underwriting; it does not complete underwriting by itself. The institution still needs to assess products, delivery obligations, settlement timing, liquidity, processing history, concentration, reserves, guarantees, and commercial terms appropriate to the relationship. These are separate institutional judgments, not outputs that should be inferred from an identity credential.

For example, a merchant selling a service months before delivery presents a different exposure pattern from a merchant delivering at the point of sale, even when both identities are equally well established. The decision record should therefore distinguish identity and authority conditions from the institution\textquotesingle{}s financial and fulfillment analysis. A refusal because an ownership document needs correction should not be described as credit default risk; an acceptable identity should not conceal an unresolved reserve requirement.

\section[\protect\toccaps{Anatomy of the Underwriting Stack}]{Anatomy of the Underwriting Stack}
The map below is a functional reference, not a claim that all acquirers use the same providers or that every provider returns only a score. Its purpose is to identify what a check can establish and where additional evidence is needed. The institution should populate a deployment-specific version with its actual providers, field coverage, update behavior, contractual rights, and exception process.

\dtabstart
\begin{longtable}{|@{\hspace{5.04bp}}P{113.28bp}@{\hspace{5.04bp}}|@{\hspace{5.28bp}}P{174.72bp}@{\hspace{5.28bp}}|@{\hspace{5.04bp}}P{179.52bp}@{\hspace{5.04bp}}|}
\hline
\hdrrow\hc{Function and source} & \hc{What the evidence can support} & \hc{What it does not establish alone}\\\hline
\hdrend
\endfirsthead
\conthead
\endhead
\textbf{Tax name/TIN combination\newline IRS TIN Matching, where eligible \cite{r3}} & Whether the submitted name/TIN pair matches under the program\textquotesingle{}s response rules. & Applicant identity, beneficial ownership, signing authority, business operations, or an EIN issuance date.\\\hline
\textbf{Identity-document attributes\newline AAMVA DLDV, where available \cite{r4}} & Field-level flags comparing submitted DL/ID information with issuing-jurisdiction data. & A full DMV record or released government photograph; the applicant\textquotesingle{}s rightful ownership of that identity; organizational authority.\\\hline
\textbf{Applicant verification\newline Evidence binding and authentication \cite{r2}} & Whether the applicant is appropriately connected to validated identity evidence, using the chosen proofing process. & Authority to represent a particular entity or execute a particular instruction.\\\hline
\textbf{Entity existence and recorded status\newline Relevant registry or qualified source} & The facts and status actually recorded by that registry, at the reported effective time. & A universal list of beneficial owners, operational legitimacy, complete ownership history, or every representative\textquotesingle{}s signing power.\\\hline
\textbf{Ownership and control\newline Disclosures plus appropriate corroboration \cite{r6}} & Identified owners and controlling persons under the applicable requirements; identity verification and evidence of disclosed relationships. & Automatic completeness of the ownership structure simply because the disclosed people are real.\\\hline
\textbf{Bank and settlement destination\newline Permissioned bank connectivity or other acceptable evidence} & Available account attributes, account access, balances, or transactions, within the actual permission and data scope. & That login access equals beneficial ownership, that all accounts are visible, or that this representative may designate the account for settlement.\\\hline
\textbf{Business model, financial capacity, and fulfillment} & Attributable processing, financial, contractual, and operational evidence used in the institution\textquotesingle{}s analysis. & Guaranteed performance, absence of future disputes, or automatic credit approval.\\\hline
\textbf{Sanctions and ownership exposure\newline Relevant lists, programs, and ownership analysis \cite{r10,r22}} & Evidence relevant to a specific sanctions evaluation, including the applicable program and matching disposition. & Universal legal clearance from a name-list result alone, or proof that an unlisted entity is never blocked.\\\hline
\textbf{PEP, adverse media, and background \cite{r11,r19}} & Specific risk indicators or records, subject to identity resolution, relevance, permissible use, and source quality. & A sanctions prohibition merely from PEP status; proven wrongdoing from an allegation; a universal eligibility verdict.\\\hline
\textbf{Network and merchant history \cite{r13,r14}} & Applicable screening, termination-history information, operating-risk indicators, and required follow-up. & A determination detached from the applicable network rules, matching review, and the institution\textquotesingle{}s facts.\\\hline
\end{longtable}
\textbf{Source precision matters.} AAMVA describes DLDV as returning match flags without releasing jurisdictional data; customers require approval. IRS matching is a purpose-limited program for eligible payers and authorized agents. An integration through a qualified intermediary may preserve source provenance, but the access path must be disclosed accurately. \cite{r3,r4} State-issued mobile driver\textquotesingle{}s licenses and other verifiable digital credentials, which FinCEN and the federal banking agencies confirmed in September 2026 may be accepted as documentary verification under the CIP rule, are expected to reduce reliance on field-matching services as issuance and acceptance spread; until then, DLDV remains the available check against issuing-jurisdiction data, and the record should identify which of the two was used. \cite{r23}

Documents remain legitimate evidence where suitable. A signed resolution or independently authenticated statement can establish facts an API does not expose. The problem is not the file format; it is an unsupported inference about authenticity, subject binding, currency, or authority. CIP permits risk-based use of documentary methods, non-documentary methods, or a combination. \cite{r5}

\section[\protect\toccaps{Failure Mode One: Distance from the Source}]{Failure Mode One: Distance from the Source}
\subsection{Define distance without inventing a risk score}
In this paper, \gbf{distance from the truth} means the gap between a decision and the evidence needed to justify it. It is not a mathematical distance, a calibrated fraud probability, or a promise that an authoritative database cannot contain an error. The institution should evaluate evidence fitness for its purpose rather than add arbitrary points for copies or intermediaries.

\dtabstart
\begin{longtable}{|@{\hspace{5.04bp}}P{144.00bp}@{\hspace{5.28bp}}|@{\hspace{5.04bp}}P{334.32bp}@{\hspace{5.04bp}}|}
\hline
\hdrrow\hc{Illustrative provenance form} & \hc{Question the institution should answer}\\\hline
\hdrend
\endfirsthead
\conthead
\endhead
\textbf{Authenticated source response} & Which attributes does the source authoritatively address, and when were they effective?\\\hline
\textbf{Qualified intermediary or maintained copy} & Can the origin, transformation, refresh behavior, and source limitations be established?\\\hline
\textbf{Applicant-presented evidence} & Can authenticity, integrity, issuer, subject relationship, and currency be validated?\\\hline
\textbf{Applicant or institutional assertion} & Who made the statement, under what responsibility, and what independent support is needed?\\\hline
\end{longtable}
These are descriptive forms, not ranked assurance levels. A digitally protected issuer credential can have stronger integrity than a fresh but unauthenticated page. A direct response can still be incomplete for the question being asked. A credible intermediary can preserve attributes and provenance without becoming their original issuer. \cite{r2}

One example shows why the access path is a question of substance rather than preference. Every merchant acquiring entity and third-party settlement organization is a payment settlement entity required to report each participating payee\textquotesingle{}s name, address, and taxpayer identification number on Form 1099-K, and payments to a payee whose TIN is missing or, on IRS notice, incorrect are subject to backup withholding. The IRS matching service exists for payers with that obligation. An acquirer that establishes a merchant\textquotesingle{}s TIN from an uploaded Form W-9 alone, when it is eligible to confirm the name and TIN pair with the IRS, has accepted applicant-presented evidence for a fact on which it already bears a federal reporting duty. The match confirms only the pair; it does not establish ownership, authority, or the date the EIN was issued. \cite{r3,r29}

\subsection{Separate validation, verification, and authority}
NIST\textquotesingle{}s identity-proofing model distinguishes validating evidence and attributes from verifying the applicant\textquotesingle{}s relationship to that evidence. Validation includes checks against authoritative or credible sources; verification establishes that the applicant is the rightful subject. Merchant authority is a further relationship: permission to act for the business within the relevant scope. A database match should not be described as proving all three. \cite{r2}

The proposed underwriting record should therefore preserve separate outcomes: evidence authentic or validated; applicant bound to the identity; organizational relationship supported; requested action within authority; and institutional risk conditions satisfied. Combining them into an unexplained \textquotedbl{}verified\textquotedbl{} flag prevents an underwriter from identifying what remains unresolved.

\subsection{Evaluate fitness across independent dimensions}
For each material assertion, record the exact subject and attribute, source authority for that attribute, authenticated retrieval path, integrity protection, relevant dates, source coverage, uncertainty, permitted use, and applicable corroboration. A field that the provider does not return must remain unavailable; it must not be inferred from an unrelated response.

A proposed \textbf{source-lineage control} records whether apparently different results share the same upstream record or applicant submission. Two vendors repeating the same error are not two independent confirmations. Corroboration should describe its actual independence rather than treating the number of responses as the number of independent observations.

\dttbox{\gbf{Preserving evidence is not the same as proving its truth.} A signature or immutable record can support attribution and detection of later alteration; it cannot make an inaccurate input accurate or keep an old fact current.}
\section[\protect\toccaps{Failure Mode Two: The Correlation Gap}]{Failure Mode Two: The Correlation Gap}
\subsection{From name matching to a relationship model}
The correlation gap consists of material relationships or inconsistencies that are not evaluated when each signal is considered alone. Closing it requires linking evidence to the same subjects, roles, business activities, destinations, and time periods. It does not require forcing every name or address to be identical.

The proposed relationship model links a verified person to a legal entity; the entity to disclosed owners or controllers; the representative to a documented role and authority; the merchant to its actual business activity and trading names; and the settlement destination to its relevant account relationship and authorized use. Each link carries its own evidence, scope, and effective period.

\dtabstart
\begin{longtable}{|@{\hspace{5.04bp}}P{133.20bp}@{\hspace{5.28bp}}|@{\hspace{5.04bp}}P{345.12bp}@{\hspace{5.04bp}}|}
\hline
\hdrrow\hc{Relationship} & \hc{Required question in the proposed application}\\\hline
\hdrend
\endfirsthead
\conthead
\endhead
\textbf{Person to entity} & Is this the correct person, and what supports the claimed relationship with this specific legal entity?\\\hline
\textbf{Entity to owner or controller} & Which ownership or control relationship is asserted, what requirements apply, and what evidence supports it?\\\hline
\textbf{Representative to action} & May this person submit information, sign terms, designate settlement instructions, or approve this change? These are different powers.\\\hline
\textbf{Entity to business activity} & Do the products, website, trading names, expected processing, and fulfillment obligations describe a coherent operation?\\\hline
\textbf{Entity to settlement destination} & Does the evidence establish the account relationship and permitted settlement arrangement, rather than mere access to a login?\\\hline
\textbf{Decision to current version} & Does the approval concern the same application, authority record, destination, and policy version being acted upon?\\\hline
\end{longtable}
\subsection{Three illustrative cases}
\gbf{A legitimate controller, different names.} A controller submits an application for a company whose executive appears on registry filings and whose legal name appears on its bank account. The names differ appropriately. The missing or supporting fact is the controller\textquotesingle{}s authority, not name equality. An independent review of the governing delegation can resolve the case without treating a legitimate organizational structure as fraud.

\gbf{Real records, wrong authority.} An applicant presents genuine business records and passes an identity check, but is authorized only to provide information. The applicant requests a new settlement destination. The proposed control requires authority for that specific change and any prescribed independent approval. The earlier onboarding result cannot be reused as permission to redirect proceeds.

\gbf{Two confirmations, one underlying assertion.} Two data providers return the same address because both ingested the same filing. The merchant claims that the address is a staffed fulfillment facility. The repeated address does not establish the operational claim. The institution requests proportionate, independent support for the facility and fulfillment arrangement.

These examples demonstrate control logic, not completed ChainIT investigations or measured fraud prevention. A discrepancy may reflect a DBA, transliteration, legitimate reorganization, delegated administration, stale records, or an error. The workflow should resolve identity and relationship questions, record alternative explanations, and permit correction before drawing a broader adverse inference.

\subsection{Authority must be established before it can be delegated}
An authority record is not self-authenticating merely because it is signed or tokenized. The institution needs an appropriate basis for accepting the person or body establishing the original organizational grant. In the ChainIT architecture, the ARP records the grant; identity, source evidence, and governance support the acceptance of that grant. A delegate cannot manufacture authority by asserting that it exists. \cite{c1}

\section[\protect\toccaps{Failure Mode Three: Point-in-Time Controls}]{Failure Mode Three: Point-in-Time Controls}
\subsection{Continuous means responsive, not omniscient}
A historical verification is evidence of what was checked at a stated time. It is not perpetual eligibility. The proposed operating model evaluates evidence at each relevant reliance event, with policy-selected freshness, source-change triggers, scheduled reviews, and escalation. It does not require every source to be queried for every login or transaction. This is consistent with the runtime model of the ChainIT Compliance Protocol. \cite{c1}

A reliance event may be initial merchant activation, acceptance of new terms, a representative change, a settlement-destination change, a new processing activity, a material exposure increase, or a periodic portfolio review. Each event should identify the evidence and authority that actually matter to the action, rather than repeat every historical check indiscriminately.

The sanctions enforcement record shows what point-in-time controls cost when the facts move. OFAC penalized Payoneer in 2021 for screening failures that included \textquotedbl{}weak algorithms that allowed close matches to SDN List entries not to be flagged\textquotedbl{} and, \textquotedbl{}during backlog periods, allowing flagged and pended payments to be automatically released without review\textquotedbl{}; Kraken in 2022 because, although it screened customer names daily, it \textquotedbl{}applied its geolocation controls only at the time of onboarding\textquotedbl{}; Poloniex in 2023 because, when it introduced a compliance program, \textquotedbl{}existing customers were not retroactively screened\textquotedbl{}; EFG International in 2024 for transactions on behalf of a client onboarded in 2009 and designated in 2014, observing that \textquotedbl{}[r]outine screening of customer information, including names and locations, as well as ongoing risk-based due diligence, is particularly important in light of frequent updates to OFAC\textquotesingle{}s sanctions programs and additions to OFAC\textquotesingle{}s SDN List\textquotedbl{}; Interactive Brokers in 2025 for, among other things, dispositioning a post-designation alert as a false positive and for a delay in obtaining ownership information relevant to the 50 Percent Rule; and BitPay in 2021, a payment processor that screened its merchants but \textquotedbl{}failed to screen location data that it obtained about its merchants\textquotesingle{} buyers.\textquotedbl{} The cases differ: Kraken and BitPay concern location controls rather than list re-screening, and only EFG and Interactive Brokers involve a customer designated after onboarding. None is a merchant-acquiring case, and none characterizes the standard as strict liability in those words. What they share is a control that was correct when applied and wrong when it mattered. \cite{r30,r31,r32,r33,r34,r35}

\subsection{Keep the clocks separate}
The proposed evidence profile records \gbf{source-effective time}, \gbf{source publication or refresh time where available}, \gbf{retrieval time}, and \gbf{decision time}, together with expiration or the next review condition. Retrieving a cached record today does not make the underlying facts current as of today. Unknown source timing is a limitation, not a zero-age value.

Policy should identify the permitted age for each evidence domain, the events requiring a new source query, and the handling of uncertain time or status. A durable identity anchor may remain useful while a bank destination, license, authority grant, or sanctions result requires new evaluation. Reauthentication proves present use of an accepted identity mechanism; it does not refresh every underlying external fact.

\dtabstart
\begin{longtable}{|@{\hspace{5.04bp}}P{138.96bp}@{\hspace{5.28bp}}|@{\hspace{5.04bp}}P{339.12bp}@{\hspace{5.28bp}}|}
\hline
\hdrrow\hc{Trigger} & \hc{Proposed controlled response}\\\hline
\hdrend
\endfirsthead
\conthead
\endhead
\textbf{Settlement destination or payout authority changes} & Bind the request to the exact destination and application version; obtain current authority, required independent approval, and relevant account evidence before activation.\\\hline
\textbf{Ownership, control, or signatory changes} & Reassess affected relationships and required evidence; do not silently transfer an earlier representative\textquotesingle{}s privileges.\\\hline
\textbf{New product, geography, activity, or material volume change} & Reassess the applicable policy, network category, financial exposure, and supporting operational evidence.\\\hline
\textbf{Source correction, designation, revocation, or material alert} & Identify dependent decisions; refresh or restrict affected uses; preserve the historical record and the new evidence separately.\\\hline
\textbf{Required source unavailable or stale} & Apply a preapproved alternative, a scoped Hold or Review, or Stale/Unknown treatment. Record the limitation and route recovery work.\\\hline
\textbf{Recovery after outage or erroneous source response} & Reconcile affected cases, refresh required evidence, release only eligible actions, and preserve the incident and reassessment trail.\\\hline
\end{longtable}
\subsection{Close the gap between approval and action}
A proposed implementation control binds an approval to the merchant, requested action, exact material instructions, relevant policy and authority versions, and a defined validity window. Immediately before the action, the receiving system checks that those inputs have not changed and that required status remains acceptable. A changed settlement destination invalidates approval for the old destination.

Version checks and one-use action identifiers can prevent concurrent updates from applying an approval to the wrong state. Delayed source events, duplicate callbacks, and replayed responses should be reconciled explicitly rather than applied in arrival order without context. A timeout after sending an instruction creates an uncertain outcome; it is not permission to send the instruction repeatedly.

A Hold is not necessarily a finding of fraud, and a source outage is not automatically grounds to suspend every merchant. The institution defines affected scope, lawful alternatives, exceptions, and escalation. A known prohibition must not be overridden merely because a provider is unavailable. These are recommended application controls, not a representation that every receiving system already enforces them.

\clearpage
\section[\protect\toccaps{What Regulators and Networks Actually Expect}]{What Regulators and Networks Actually Expect}
The legal and contractual analysis begins with the entity and the activity. A sponsor bank, acquiring bank, payment facilitator, independent sales organization, and technology provider do not automatically have identical direct obligations. The map below distinguishes rules and supervisory guidance from contractual allocation and proposed implementation practice. It is not a finding that any particular deployment satisfies the listed requirements.

\dtabstart
\begin{longtable}{|@{\hspace{5.04bp}}P{145.68bp}@{\hspace{5.04bp}}|@{\hspace{5.28bp}}P{332.64bp}@{\hspace{5.04bp}}|}
\hline
\hdrrow\hc{Authority or framework} & \hc{Relevant boundary and architectural implication}\\\hline
\hdrend
\endfirsthead
\conthead
\endhead
\textbf{Bank CIP\newline 31 C.F.R. 1020.220 \cite{r5}} & Risk-based identity verification may use documents, non-documentary methods, or both. Preserve methods and outcomes; do not claim that direct-source technology is mandated by name.\\\hline
\textbf{Beneficial ownership\newline 31 C.F.R. 1010.230 and FinCEN relief \cite{r6,r8}} & Apply the covered-institution, legal-entity-customer, ownership/control, and exception rules. Distinguish identifying owners and verifying identities from independently proving every ownership assertion.\\\hline
\textbf{Ongoing bank CDD\newline 31 C.F.R. 1020.210(a)(2)(v) \cite{r7}} & Nature and purpose, risk profiles, and ongoing monitoring/risk-based updating belong to the bank\textquotesingle{}s program. This is distinct from the beneficial-ownership provision.\\\hline
\textbf{OFAC sanctions \cite{r10,r22}} & Assess the applicable programs, parties, ownership, and transaction context. A name-list result is evidence, not a universal sanctions clearance.\\\hline
\textbf{Third-party risk guidance \cite{r12}} & Outsourcing does not remove the bank\textquotesingle{}s responsibility for its activities. Address diligence, contracts, monitoring, concentration, contingency, and termination. On September 15, 2026 the agencies proposed to rescind and replace the 2023 guidance; it remains in effect pending that rulemaking, and the proposal restates the retained-responsibility principle. \cite{r12,r24}\\\hline
\textbf{Card-network standards \cite{r13,r14}} & Apply the operative regional and product-specific screening, monitoring, record, and reporting requirements through the appropriate acquirer/PayFac relationships.\\\hline
\textbf{Nacha Operating Rules \cite{r15}} & For ACH participants in scope, incorporate applicable risk-based fraud-monitoring processes. An ACH obligation is not established merely by describing a generic payment platform.\\\hline
\textbf{Tax reporting and TIN matching \cite{r3}} & Determine the actual reporting party, eligible user or authorized-agent role, and permitted purpose. Pre-filing mismatch and backup-withholding procedures are not interchangeable.\\\hline
\textbf{Consumer reports and business credit \cite{r19,r20}} & Determine actual FCRA roles and permissible use. Where the product is credit, apply the relevant Regulation B requirements, including business-credit distinctions.\\\hline
\textbf{Source access and confidential information \cite{r17,r21}} & Validate lawful access, permissions, retention, and disclosure. Consent alone does not override program restrictions or SAR confidentiality.\\\hline
\end{longtable}
\subsection{Current-law distinctions relevant to this edition}
\textbf{CDD reuse.} FinCEN\textquotesingle{}s February 13, 2026 exceptive relief permits covered institutions to limit repeated beneficial-owner identification and verification to the first account, circumstances calling prior information into question, and risk-based ongoing procedures, rather than repeating the exercise at every new account. This supports governed reuse, not permanent reliance regardless of changed facts. \cite{r6,r8}

\gbf{CTA reporting.} Treasury announced a final BOI reporting rule on August 11, 2026, effective August 14, 2026. The rule permanently exempts entities formed in the United States (formerly domestic reporting companies) from reporting, and exempts U.S.-person beneficial owners and company applicants from being reported; entities formed under foreign law and registered to do business in the United States remain reporting companies. CTA reporting and financial-institution CDD are separate regimes. A change in domestic BOI reporting does not, by itself, create a new direct CDD duty for every acquirer or make a registry a complete ownership source. \cite{r9}

\gbf{Model governance.} The Federal Reserve\textquotesingle{}s April 17, 2026 SR 26-2 supersedes SR 11-7. Governance should reflect the institution, use, materiality, and actual model characteristics. A deterministic rule should not automatically be described as a quantitative model; models used to generate risk inputs need their own appropriate validation and monitoring. \cite{r16}

\gbf{Account-data access}\textbf{.} The CFPB\textquotesingle{}s personal financial data rights resource reports that the Section 1033 compliance dates were stayed by a court on October 29, 2025. The architecture should rely on the actual authorized connection and permitted use, not an assumption that the rule creates unrestricted access to all merchant accounts. \cite{r17}

\gbf{Proposals are not operative requirements}\textbf{.} FinCEN\textquotesingle{}s April 7, 2026 AML/CFT program proposal replaced and withdrew its July 2024 proposal. Proposed requirements are not treated as binding duties in this paper. Institutions should maintain a versioned legal-obligation inventory rather than embed an obsolete proposal in production policy. \cite{r18}

\subsection{Translate network rules into evidence, not slogans}
Visa\textquotesingle{}s April 18, 2026 public rules include pre-contract merchant due diligence in section 5.2.1.2. The Mastercard Security Rules and Procedures, Merchant Edition (public edition of August 4, 2026), address required screening, retention of investigative records, and ongoing monitoring in sections 7.1 and 7.2, including reassessment on specified material changes. These obligations support linked, retained evidence and event-driven workflows; they do not establish that one named vendor or technical design is sufficient. \cite{r13,r14}

Nacha\textquotesingle{}s Phase 2 fraud-monitoring changes took effect June 19, 2026, with June 22 identified as the practical implementation date because of the banking holiday. The changes extend relevant risk-based monitoring requirements to remaining participants in scope. They should be mapped to the actual ACH role and flow, rather than paraphrased as a requirement to use a particular product or to make every check in real time. \cite{r15}

Public rulebooks are reference snapshots. Network bulletins, regional variations, contracts, and the institution\textquotesingle{}s controlling edition must be reconciled before implementation. A useful policy record identifies the rule source, effective date, accountable owner, covered activity, evidence requirement, and review schedule. The technology then implements that approved mapping.

\subsection{Do not collapse unlike compliance concepts}
The CDD ownership prong generally uses a 25-percent threshold, together with a separate control prong and applicable exceptions. OFAC\textquotesingle{}s 50 Percent Rule addresses ownership by blocked persons, including relevant aggregate direct or indirect ownership. These are different tests. An entity can require sanctions analysis even when it is not itself named on a list. \cite{r6,r10}

PEP status is not itself an OFAC prohibition. The U.S. interagency statement calls for risk-based CDD appropriate to the relationship rather than assuming that every PEP requires a categorical outcome. Similarly, an unresolved media allegation, a case-review state, and a confirmed legal restriction should remain distinct evidence and decision categories. \cite{r11}

\section[\protect\toccaps{A Different Architecture: Connected, Current, Attributable}]{A Different Architecture: Connected, Current, Attributable}
\subsection{The ChainIT role}
This paper proposes a merchant-underwriting application of the \gbf{ChainIT \href{https://chainit.com/white-papers/}{Transaction Truth} Architecture}. It is not a separate underwriting protocol or a replacement for the institution\textquotesingle{}s underwriting authority. The ChainIT Compliance Protocol supplies purpose-specific evidence and compliance decisions. The ChainIT \href{https://chainit.com/insights/chainit-launches-provable-authority-as-u-s-generative-ai-enabled-fraud-losses-are-projected-to-reach-40-billion-by-2027/}{Authority Protocol} separately evaluates authority for the applicable session or exact transaction. \cite{c1}

\href{https://chainit.com/product/identity-verification/}{ChainIT ID} and \href{https://chainit.com/product/business-verification/}{ChainIT Org ID} provide the individual and legal-entity identity anchors. The ARP records organizational authority. \href{https://www.pactvera.com/}{Pactvera} connects relevant agreement and transaction conditions; \href{https://chainit.com/product/data-tokenization/}{VDTs} preserve discrete assertions or lifecycle events; \href{https://chainit.com/}{Valitorum} seals a terminal evidence record. The identity and Pactvera application materials provide related technical foundations, not proof that every proposed underwriting feature is deployed or covered by a particular patent claim. [\citen{c1}-\citen{c3}]

\dtabstart
\begin{longtable}{|@{\hspace{5.04bp}}P{120.96bp}@{\hspace{5.28bp}}|@{\hspace{5.04bp}}P{357.12bp}@{\hspace{5.28bp}}|}
\hline
\hdrrow\hc{Layer} & \hc{Role in the proposed merchant application}\\\hline
\hdrend
\endfirsthead
\conthead
\endhead
\textbf{Verified subjects} & Identify the applicant, legal entity, relevant owners or controllers, and authorized representatives without treating them as interchangeable.\\\hline
\textbf{Relationship and authority} & Establish and maintain supported links; resolve the governing ARP and the scope needed for submission, signature, settlement designation, and changes.\\\hline
\textbf{Evidence Assertions} & Represent sourced observations separately, preserving source class, identity, timestamps, limits, and permitted disclosure.\\\hline
\textbf{Policy and predicates} & A versioned CPP selects required evidence and conditions. The BRE returns deterministic, reason-coded true/false predicate results, not a risk score or disposition.\\\hline
\textbf{Institution-configured workflow} & Compose predicate results, review and control completion, and the relevant disposition. Keep financial and commercial acceptance separately attributable.\\\hline
\textbf{Action boundary} & Check the exact instruction, applicable final decision, authority, freshness, and receiving-system enforcement before the consequential action.\\\hline
\textbf{Lifecycle evidence} & Link decision, action, authenticated outcome, correction, and later reassessment. Seal terminal records without altering earlier evidence.\\\hline
\end{longtable}
\gbf{Control view:} Sources and permitted assertions feed a subject-and-relationship record. Policy selects required checks. The BRE returns predicates. The institution-configured Workflow Engine produces a disposition. Authority and action-specific requirements govern what may proceed. Outcome evidence returns to the same logically linked record. This is an application-level control view of C1, not a claim that source responses themselves authorize execution.

\subsection{A logically unified evidence record}
Logical unity means that an authorized reviewer can follow the relevant facts and decisions without requiring unrestricted access to every raw payload. Preserve the provenance and integrity references needed for reconstruction; retain or retrieve source data only under the appropriate rights and safeguards. Record when retention rights prevent keeping the underlying source payload, and validate the institution\textquotesingle{}s permitted reconstruction method.

The source classes remain \gbf{authoritative-source evidence, qualified third-party evidence, and institution-generated evidence}. A ChainIT signature over an authenticated API response is a ChainIT attestation about that response, not a government signature. An institution-generated conclusion is not silently promoted to an issuer fact. Each assertion remains distinct from the institution\textquotesingle{}s final decision. \cite{c1}

Sensitive identity and bank records should not be placed on a publicly readable ledger as an underwriting convenience. The proposed profile uses controlled storage and references or appropriate cryptographic commitments. A hash is not automatically anonymous: low-entropy or linkable personal data requires appropriate privacy design. The record should also distinguish expiry of permission, correction of an assertion, and retention or deletion of source data.

\subsection{Proposed operating sequence}
\dtabstart
\begin{longtable}{|@{\hspace{5.04bp}}P{120.96bp}@{\hspace{5.28bp}}|@{\hspace{5.04bp}}P{357.12bp}@{\hspace{5.28bp}}|}
\hline
\hdrrow\hc{Stage} & \hc{Required result before the next consequential step}\\\hline
\hdrend
\endfirsthead
\conthead
\endhead
\textbf{1. Define the reliance event} & Specify merchant, action, legal entity, institution, requested terms or destination, policy context, and authorized data uses.\\\hline
\textbf{2. Verify subjects and relationships} & Establish identity and entity facts; resolve the relevant relationship and original or delegated authority.\\\hline
\textbf{3. Assemble fit-for-purpose evidence} & Acquire or reuse permitted evidence within policy; label unsupported fields, shared lineage, uncertainty, and stale or unavailable sources.\\\hline
\textbf{4. Evaluate and review} & Run reason-coded predicates; compose the institution\textquotesingle{}s workflow disposition; complete controls and separately record financial acceptance.\\\hline
\textbf{5. Recheck and act} & Confirm material instruction/version stability and current permission at the receiving boundary. A conditional disposition is not final approval.\\\hline
\textbf{6. Record and monitor} & Record the authenticated action outcome, seal a terminal record when appropriate, and route changes or exceptions to fresh reliance events.\\\hline
\end{longtable}
For a \gbf{payment transaction}, the established protocol sequence is more specific: final transaction Allow follows completion of required controls and approvals; available capacity is reserved; a short-lived, single-use EAC is issued for the exact transaction; and a supported payment boundary consumes it. A SAC permits defined session activity but cannot authorize payment. Merchant activation is not payment authorization, and an old decision receipt is not a reusable credential. \cite{c1}

Where \href{https://chainit.com/product/digital-pay/}{ChainIT Pay} or a compatible partner adapter is used, the implementation must demonstrate its actual enforcement boundary and outcome handling. The architecture does not imply that a downstream bank or processor accepts a credential merely because ChainIT can generate it. A recorded request, an accepted instruction, and completed settlement remain separate events.

\subsection{Architecture and implementation status}
The classifications below preserve the public status distinctions in \href{https://chainit.com/white-papers/}{Provable Compliance} v4.1. They describe foundations and development status, not a new independent deployment certification. The merchant-specific application profile in this paper is proposed. Availability and coverage must be established for the contracted deployment. \cite{c1}

\dtabstart
\begin{longtable}{|@{\hspace{5.04bp}}P{208.80bp}@{\hspace{5.28bp}}|@{\hspace{5.04bp}}P{269.52bp}@{\hspace{5.04bp}}|}
\hline
\hdrrow\hc{Components} & \hc{Publication classification}\\\hline
\hdrend
\endfirsthead
\conthead
\endhead
\textbf{\href{https://chainit.com/product/identity-verification/}{ChainIT ID}, \href{https://chainit.com/product/business-verification/}{ChainIT Org ID}, ARP; \href{https://www.pactvera.com/}{Pactvera}, \href{https://chainit.com/}{Valitorum}; BRE predicate foundation} & \textbf{Current platform foundation.} Specific assurance, source access, lifecycle configuration, and predicates remain deployment-specific.\\\hline
\textbf{Evidence Assertion \href{https://chainit.com/product/data-tokenization/}{VDTs}; CPP; status/freshness; compliance Workflow Engine; Compliance Decision VDT} & \textbf{Development / pilot.} Being built, integrated, or validated; not represented as generally available in every environment.\\\hline
\textbf{Disclosure-tier controls; Supplemental \href{https://chainit.com/}{Valitorum}} & \textbf{Development / pilot.} Validate permitted audiences, exception handling, and retained evidence for the particular implementation.\\\hline
\textbf{Compliance Decision Permits interface; formal manual-review/override architecture} & \textbf{Proposed protocol control.} Defined controls are not represented as universally operational.\\\hline
\textbf{External evidence sources and partner rail-adapter enforcement} & \textbf{External dependency.} Access, fields, permitted uses, availability, and enforcement depend on source or partner arrangements.\\\hline
\textbf{Merchant application profile, change-control tests, and pilot metrics proposed here} & \textbf{Proposed application design.} Requires integration, validation, institutional approval, and implementation-specific evidence.\\\hline
\end{longtable}
No statement in this paper should be read as asserting direct production access to every IRS, DMV, registry, bureau, or bank; complete replacement of every screening provider; a particular independent assurance report; or a quantified reduction in fraud. Those are separate, testable capability or performance claims.

\section[\protect\toccaps{The Objections a Careful Buyer Will Raise}]{The Objections a Careful Buyer Will Raise}
\subsection{Does consolidation create a single point of failure?}
It can. A unified record does not require one upstream source or one operational dependency. The proposed procurement test is to map shared providers, shared source systems, control-plane failure, and data-export dependencies. Two vendor contracts backed by the same source are not necessarily independent fallbacks. Test continuity, recovery, evidence reconstruction, and exit before relying on consolidation. This is consistent with lifecycle-based third-party risk management. \cite{r12}

Fallbacks should preserve their actual assurance. Where a required source is unavailable, a weaker substitute must not silently receive the original source\textquotesingle{}s label. Contractual service levels should distinguish platform availability, source availability, evidence freshness, and time to resolve an exception. An impressive uptime figure does not demonstrate that required evidence was current.

\subsection{What about data rights, privacy, and confidentiality?}
Data access must be lawful for the specific party, purpose, subject, and use. Applicant consent is important where required, but it is not a universal override of source contracts, tax-program restrictions, consumer-report requirements, or financial privacy obligations. Allocate controller, service-provider, report-user, and other relevant roles to the actual arrangement, with counsel resolving uncertain legal characterization. \cite{r3,r17,r19}

Use audience-specific disclosure. A merchant may receive the reason it must correct an account or authority record, while a sponsor-bank reviewer receives additional permitted evidence. Source documents, biometric templates, full account identifiers, and investigative case details should be available only to appropriate audiences. The institution should define retention, legal holds, correction, export, and deletion schedules rather than retain everything indefinitely.

SAR confidentiality requires particular care. FinCEN\textquotesingle{}s guidance addresses information revealing the existence or non-existence of a SAR, not just disclosure of the form itself. Public credentials, customer-facing reason codes, timing responses, and shared evidence packages must not expose protected SAR information. A generic \textquotedbl{}AML clear\textquotedbl{} credential should not be used to represent confidential reporting status. \cite{r21}

\subsection{Can an applicant challenge a bad record?}
The proposed workflow includes a correction and review path, an accountable reviewer, and a service target for unresolved cases. Preserve the original source response and decision; append corrected evidence, the reason for correction, and the new disposition. Do not rewrite history to make the original decision appear better supported. Restrict current reliance on an assertion that has been superseded or credibly disputed.

Handle thin-file businesses, naming differences, legitimate registered-agent addresses, and nontraditional operating models proportionately. Missing evidence is not automatically fraud. The institution should define acceptable alternatives and review for unjustified exclusion. Where consumer-report or credit rules apply, the required notice and dispute processes remain distinct from a generic product-support ticket. \cite{r19,r20}

\subsection{What is the role of models, AI, and agents?}
The proposed design permits AI to assist with document extraction, evidence organization, and case summaries, provided its outputs remain attributable to the underlying material and its uncertain interpretations are visible. An AI-generated explanation is not new source evidence. Approved policy, qualified human review where required, and deterministic action gates must control consequential decisions.

Models producing risk inputs should have suitable documentation, validation, monitoring, and change control. The BRE\textquotesingle{}s deterministic predicates and the Workflow Engine\textquotesingle{}s disposition remain distinct from those models. An automated agent may submit an application or propose an action only under an accepted principal and bounded authority; it may not grant itself authority or bypass required approval. \cite{r16,c1}

\subsection{Must the buyer replace the existing stack?}
No. Begin with a bounded segment or a high-impact change such as settlement-destination updates. Run evidence and decisions in parallel. Retain incumbent controls while testing interoperability, duplicate handling, exception routing, and source rights. Replace or consolidate a function only after demonstrating equivalent or better controls and an acceptable operating cost. Procurement savings should be a measured consequence, not the premise.

\clearpage
\section[\protect\toccaps{Measuring the Difference}]{Measuring the Difference}
\subsection{A proposed pilot, not a claim of demonstrated performance}
The pilot is designed to answer three questions: whether the connected evidence record is more complete than the incumbent process, whether legitimate merchants experience less friction, and whether adjudicated disagreements between the two favor the connected record. The safeguards that follow exist so that the answers can be believed. Enroll a predefined merchant segment for 90 days. Operate ChainIT in shadow mode alongside the incumbent process, with material safety alerts handled under the institution\textquotesingle{}s approved escalation policy. Freeze the comparison definitions, policy versions, exclusion criteria, and primary metrics before evaluating results. Maintain an auditable change log for necessary mid-pilot revisions.

Stratify by relevant business model, merchant size, new versus established business, geography, and source coverage. Include incomplete applications and source failures in the appropriate denominators rather than removing inconvenient cases. Record elapsed time from both application start and completion so faster decisions are not manufactured by starting the clock late.

\subsection{Adjudicate disagreement and preserve the counterfactual limit}
Independent reviewers, blinded to the producing system where practical, should classify disagreements as confirmed fraud, legitimate application, unresolved suspicion, data or matching error, or policy difference. A match requiring review is not fraud. When evidence remains insufficient, retain the unresolved label rather than converting uncertainty into a favorable performance claim.

A shadow decision does not create a second observed financial outcome. For applications declined by the operating system, the pilot generally cannot observe what their processed loss experience would have been. Downstream loss comparisons must identify the decisions actually implemented, follow-up completeness, selection effects, and changes in controls or merchant mix. A causal avoided-loss claim requires a design supporting that inference, not just a list of rejected applications.

Observe eligible activated merchants for the required 90- and 180-day windows after activation. A merchant enrolled near the end of a 90-day enrollment period does not provide 180 days of outcome evidence by the enrollment close. Publish the enrollment window, follow-up window, sample size, censoring, and adjudication rules with any results.

\dtabstart
\begin{longtable}{|@{\hspace{5.04bp}}P{128.16bp}@{\hspace{5.28bp}}|@{\hspace{5.04bp}}P{349.92bp}@{\hspace{5.28bp}}|}
\hline
\hdrrow\hc{Measure} & \hc{Definition for the proposed pilot}\\\hline
\hdrend
\endfirsthead
\conthead
\endhead
\textbf{Evidence completeness} & Decisions with all policy-required evidence or a documented permissible alternative, divided by all decisions requiring that evidence. Report missing and unavailable categories separately.\\\hline
\textbf{Adjudicated detection and false positives} & Report confirmed-fraud detection among adjudicated fraud cases, and fraud-related flags among adjudicated legitimate cases. Also report unresolved cases and the fraction adjudicated.\\\hline
\textbf{Manual review and merchant friction} & Manual-review cases divided by all evaluated applications; completion/abandonment among started applications; median and 95th-percentile time for comparable cohorts.\\\hline
\textbf{Freshness and source failure treatment} & Events handled according to approved stale/unavailable-evidence policy divided by events to which that policy applied; report source age and latency distributions.\\\hline
\textbf{Authority and instruction integrity} & Number of unauthorized, expired, replayed, cross-subject, or materially changed instructions admitted in defined control tests and observed operations.\\\hline
\textbf{Financial outcomes} & Losses, chargebacks, refunds, and other agreed outcomes for actual activated cohorts, with explicit amount/volume denominators and 90/180-day follow-up.\\\hline
\textbf{Fully loaded operating cost} & Provider fees, repeat queries, integration/operations allocation, review labor, incident and exception handling, and governance cost divided by the stated activity denominator.\\\hline
\textbf{Evidence reconstruction} & Proportion of sampled decisions reconstructable under authorized access, plus median and 95th-percentile retrieval/reconstruction time.\\\hline
\end{longtable}
\subsection{Test control failures deliberately}
The proposed test suite includes wrong-subject and wrong-tenant evidence; common-source duplicate corroboration; source-response replay; expired authority; destination changes after approval; simultaneous approval and revocation; duplicate callbacks; missing source-effective times; source outages; inappropriate manual overrides; protected-information leakage; and an uncertain downstream execution outcome.

The acceptance target for defined critical tests is zero admitted unauthorized or materially altered actions, zero acceptance of replayed execution permission, and zero protected-information disclosures. Also require correct recording and escalation of every injected failure. A zero count in a finite test suite is evidence about those tests, not proof of zero real-world risk. Record the tested coverage and residual risks.

\subsection{Decide whether to scale}
Set quantitative thresholds with the sponsor and operating institution before the pilot. Scale only when control results, merchant outcomes, operating burden, data rights, source coverage, and evidence reconstruction meet those agreed thresholds. A faster approval time should not compensate for a critical authorization defect. A higher alert rate should not be called better detection without adjudication.

The commercial evaluation should report verified changes in cost and workflow separately from projected benefits. Subtract migration and ongoing governance costs when assessing savings. Do not monetize every suspicious decline at its requested processing volume or treat potential exposure as realized avoided loss.

\clearpage
\section[\protect\toccaps{Self-Assessment: Ten Questions for Your Current Stack}]{Self-Assessment: Ten Questions for Your Current Stack}
For each question, record \gbf{Demonstrated, Partly demonstrated, or Not demonstrated}, together with an evidence reference and a responsible owner. These responses are a control inventory, not a calibrated vendor score.

\qhead{1. What does each check actually establish?}
Can you name the source, exact supported attribute, access path, limitations, and permitted use rather than only the provider or score?

\qhead{2. Is the applicant bound to the identity?}
Can you distinguish evidence validation, applicant verification, organizational relationship, and authority for the requested action?

\qhead{3. Are the relationships coherent?}
Can you explain the connections among legal entity, owners/controllers, representatives, trading activity, and settlement destination without relying on name equality?

\qhead{4. Are confirmations genuinely independent?}
Can you identify shared upstream records, copies, or applicant submissions behind apparently separate provider responses?

\qhead{5. Is evidence current enough for this event?}
Do you preserve source-effective and retrieval times, apply domain-specific freshness, and expose unknown or stale state?

\qhead{6. What prevents an unauthorized change?}
Does approval bind the exact settlement destination or other material instruction, with current authority and a final version check?

\qhead{7. What happens when a source or system fails?}
Is there a tested, policy-conforming fallback or scoped hold, including reconciliation after uncertain outcomes?

\qhead{8. Can you reconstruct one decision?}
Can an authorized reviewer trace evidence, policy, predicates, disposition, approvals, exceptions, and action outcome within an agreed service target?

\qhead{9. Can you protect and correct the record?}
Are data rights, restricted audiences, SAR confidentiality, correction, retention, and current-status changes enforced without rewriting history?

\qhead{10. Can you prove the business result?}
Do pilot measures include adjudicated errors, incomplete applications, friction, actual loss outcomes, full cost, and known limitations?

\clearpage
\section[\protect\toccaps{Related Work}]{Related Work}\label{sec:related}
\textbf{Source dependence and corroboration.} Truth-discovery methods estimate the reliability of sources that report conflicting values~\cite{r37}, and work on source dependence shows that sources which copy one another should not be counted as independent support for a value~\cite{r36}. The source-lineage control in Section~3.3 and the case of two confirmations from one underlying assertion in Section~4.2 apply this principle to underwriting evidence.

\textbf{Record linkage and relationship evidence.} Probabilistic record linkage decides whether two records refer to the same entity~\cite{r38}. The relationship model in Section~4 does not require forcing every name or address to be identical; each link between person, entity, owner or controller, representative, business activity, and settlement destination carries its own evidence, scope, and effective period.

\textbf{Provenance, credentials, and tamper evidence.} The W3C PROV data model describes the entities, activities, and agents involved in producing data~\cite{r39}; verifiable credentials let a holder present issuer-signed claims that a verifier can check cryptographically~\cite{r41}; and tamper-evident logs let auditors detect later alteration of recorded events~\cite{r40}. As Section~3 states, such mechanisms support attribution and detection of later alteration, but they cannot make an inaccurate input accurate or keep an old fact current.

\textbf{Point-in-time checks and fraud scoring.} Time-of-check-to-time-of-use flaws arise when a condition verified earlier no longer holds when it is relied on~\cite{r42}; Section~5.3 addresses the corresponding gap between approval and action. Statistical fraud detection is reviewed in~\cite{r43}. This paper does not propose a new risk score (Section~3.1) and treats model outputs as inputs that need their own validation and monitoring (Sections~6.1 and~8.4).

\unnumbered{Conclusion: From Checks to Defensible Decisions}
Merchant underwriting does not become stronger merely because another check is added or another vendor is removed. It becomes stronger when the institution can explain what was established, how the evidence applies to the merchant and its representatives, why the proposed action is permitted, and what must happen when those conditions change.

Provenance reduces unsupported assumptions about sources. Correlation connects identities, organizational relationships, authority, business activity, and settlement instructions. Persistence keeps the decision responsive to new facts and new reliance events. Together, these principles support a record that is useful before an action, enforceable at the relevant control boundary, and reviewable afterward.

The proposed ChainIT merchant application brings these principles into the \href{https://chainit.com/white-papers/}{Transaction Truth} Architecture without replacing institutional judgment, lawful access requirements, sponsor-bank oversight, or network obligations. The next practical step is a scoped evidence-and-control assessment followed by the defined shadow pilot: one merchant segment, explicit owners, measurable outcomes, and a documented scale decision.

\dttbox{\gbf{Verify the evidence. Establish the relationship. Resolve the authority. Evaluate the present action. Preserve what occurred.}}
\gbf{Trust Nothing. Verify Everything. With ChainIT.}

\clearpage
\unnumbered{Appendix A. Proposed Evidence and Decision Receipt}
This proposed application profile supports reconstruction of a merchant decision. It is not a production API contract, new token standard, or execution credential. Field names, storage, access, and disclosure remain implementation-specific. \cite{c1}

{\setlength{\parskip}{4bp}\vspace{2.5bp}
\gbf{Identity and context.} Institution/tenant; merchant legal-entity reference; applicant/representative reference; event identifier and type; application or instruction version; audience and purpose.

\gbf{Evidence assertions.} Assertion identifiers; source classes; actual source/intermediary; precise attributes; lineage; integrity/verification method; source response status; supported and unavailable fields.

\gbf{Time and current state.} Effective/as-of time; publication or source refresh where available; retrieval time; decision time; expiry/review condition; revocation/supersession/status references; clock limitations.

\gbf{Relationships and authority.} Relevant relationship evidence; ARP or delegation reference/version; permitted action and limits; original-grant acceptance basis; required approvals and completion evidence.

\gbf{Policy and evaluation.} CPP and other applicable policy versions; reason-coded predicates; institution-configured disposition; unresolved limitations; separate financial/commercial assessment reference.

\gbf{Controls and review.} Required controls; reviewer role and permission; review rationale; override scope and expiry where permitted; final re-evaluation; original and amended decision linkage.

\gbf{Action and outcome.} Exact material instruction reference; expected version; action identifier; receiving-system acknowledgement; known, pending, failed, or uncertain outcome; reconciliation reference.

\gbf{Integrity and disclosure}\textbf{.} Issuer and verification method; signed content scope; permitted audience; retention/legal-hold class; controlled source references; export/reconstruction instructions; terminal evidence linkage.\par}
\vspace{5.85bp}
\subsection*{\color{dgreen}Illustrative receipt: change of settlement destination}
Fictional example. Controller P-18 requests a settlement-destination change for Merchant M-204 from account reference A-17 to A-29 under ARP-12 version 4. The policy requires current account evidence and a second authorized approver. All identifiers are fictional.

Allow with Controls does not activate A-29. After the evidence and independent approval are complete, the workflow re-evaluates the request. Final Allow binds A-29, request version 8, current authority, and a limited validity window.

The receiving system rechecks that exact instruction and required status. Substitution of A-30 invalidates the earlier approval. A timeout creates an uncertain outcome requiring reconciliation before retry. Confirmation proves the destination update, not a later payment or settlement.

The receipt preserves history; it cannot authorize another change. Its customer-facing view excludes unnecessary personal data, source secrets, investigative details, and SAR status.

\clearpage
\unnumbered{Definitions and Abbreviations}
The following ChainIT terms, regulators, laws, standards, and abbreviations are used in this paper. ChainIT protocol terms follow the definitions in \href{https://chainit.com/white-papers/}{Provable Compliance}, CIT-PC-WP-4.1. \cite{c1}

\dtabstart
\begin{longtable}{|@{\hspace{5.04bp}}P{130.32bp}@{\hspace{5.28bp}}|@{\hspace{5.04bp}}P{347.76bp}@{\hspace{5.28bp}}|}
\hline
\hdrrow\hc{Term} & \hc{Meaning in this paper}\\\hline
\hdrend
\endfirsthead
\conthead
\endhead
\textbf{Acquirer / payment facilitator} & Participants performing the relevant merchant-acquiring or sponsored-merchant functions under their actual legal and network arrangements.\\\hline
\textbf{Authoritative source} & A source with authority over the particular fact at issue; not a universal source of truth for every underwriting question.\\\hline
\textbf{Evidence validation / applicant verification} & Validation concerns the evidence and attributes; applicant verification connects the applicant to validated identity evidence. \cite{r2}\\\hline
\textbf{Reliance event} & A defined moment at which evidence is used for a consequential decision, such as activation, a change request, a review, or a transaction.\\\hline
\textbf{VDT} & \href{https://chainit.com/product/data-tokenization/}{Validated Data Token}: the ChainIT evidence-artifact family. An Evidence Assertion VDT represents a sourced observation or institutional state, not a universal conclusion. \cite{c1}\\\hline
\textbf{ARP} & Authority Resolution \href{https://www.pactvera.com/}{Pactvera}: the canonical, versioned organizational authority record. \cite{c1}\\\hline
\textbf{CPP / APP} & Compliance Policy Profile / Authority Policy Profile. Policies select evaluation conditions; they do not create the organizational grant. \cite{c1}\\\hline
\textbf{BRE / Workflow Engine} & The Business Rules Engine returns deterministic reason-coded predicates. The institution-configured Workflow Engine composes the relevant disposition. \cite{c1}\\\hline
\textbf{Compliance Decision Permits} & The compliance-side predicate consumed by the separate \href{https://chainit.com/insights/chainit-launches-provable-authority-as-u-s-generative-ai-enabled-fraud-losses-are-projected-to-reach-40-billion-by-2027/}{Authority Protocol}. It is not independent payment authority. \cite{c1}\\\hline
\textbf{SAC / EAC} & Session Authorization Credential / Execution Authorization Credential. A SAC permits defined session activity; an EAC authorizes a single exact transaction under the applicable protocol. \cite{c1}\\\hline
\textbf{\href{https://www.pactvera.com/}{Pactvera} / \href{https://chainit.com/}{Valitorum}} & The commercial transaction framework / the sealed terminal evidence record. A Supplemental \href{https://chainit.com/}{Valitorum} records later exceptions without modifying the original. \cite{c1}\\\hline
\textbf{Stale/Unknown} & Evidence or status cannot satisfy the applicable currency or certainty requirement. It is not, by itself, a fraud finding. \cite{c1}\\\hline
\textbf{TOCTOU} & Time-of-check to time-of-use: the risk that material facts or instructions change between evaluation and the action relying on it.\\\hline
\textbf{KYC / KYB / CDD} & Know your customer / know your business / customer due diligence. Their operational and legal scope depends on the actual institution and activity.\\\hline
\textbf{PEP / SAR / MCC} & Politically exposed person / suspicious activity report / merchant category code. These are different risk, reporting, and classification concepts.\\\hline
\textbf{Regulators and bodies} & \textbf{Meaning in this paper}\\\hline
\textbf{FinCEN} & Financial Crimes Enforcement Network, the bureau of the U.S. Department of the Treasury that administers the Bank Secrecy Act.\\\hline
\textbf{OFAC} & Office of Foreign Assets Control, the Treasury office that administers and enforces U.S. economic sanctions and maintains the SDN List.\\\hline
\textbf{Federal Reserve; SR letter} & The Federal Reserve System and its Board of Governors, the federal regulator of state member banks and bank holding companies; SR letters are the Board\textquotesingle{}s Supervision and Regulation letters.\\\hline
\textbf{OCC; FDIC; NCUA} & Office of the Comptroller of the Currency (national banks and federal savings associations); Federal Deposit Insurance Corporation (state nonmember banks); National Credit Union Administration (federally insured credit unions). With the Federal Reserve, the federal banking agencies.\\\hline
\textbf{FFIEC} & Federal Financial Institutions Examination Council, the interagency body that publishes the BSA/AML Examination Manual used by federal examiners.\\\hline
\textbf{CFPB} & Consumer Financial Protection Bureau, which administers Regulation B, Regulation P, and the Section 1033 personal financial data rights rule.\\\hline
\textbf{FTC} & Federal Trade Commission, the federal consumer protection agency; it operates the Consumer Sentinel Network of consumer fraud reports and publishes FCRA business guidance.\\\hline
\textbf{IRS} & Internal Revenue Service, which operates the TIN Matching program and receives Form 1099-K information returns.\\\hline
\textbf{AAMVA; DMV} & American Association of Motor Vehicle Administrators, the association of state motor vehicle agencies (DMVs) that operates the DLDV service.\\\hline
\textbf{NIST} & National Institute of Standards and Technology, the Commerce Department agency that publishes the SP 800-63 Digital Identity Guidelines.\\\hline
\textbf{Nacha} & The organization that administers the ACH network and publishes the Nacha Operating Rules.\\\hline
\textbf{Laws, rules, and standards} & \textbf{Meaning in this paper}\\\hline
\textbf{BSA; AML/CFT} & Bank Secrecy Act, 31 U.S.C. 5311 et seq., the principal U.S. anti-money laundering statute; anti-money laundering and countering the financing of terrorism, the program, due diligence, monitoring, and reporting obligations it imposes.\\\hline
\textbf{CIP} & Customer Identification Program, the BSA requirement (for banks, 31 C.F.R. 1020.220) to verify the identity of each customer using documentary or non-documentary methods.\\\hline
\textbf{CDD Rule; ongoing CDD} & FinCEN\textquotesingle{}s 2016 Customer Due Diligence rule, whose beneficial-ownership provision is 31 C.F.R. 1010.230 and whose nature-and-purpose and ongoing-monitoring elements are codified in the AML program rules, for banks at 31 C.F.R. 1020.210(a)(2)(v).\\\hline
\textbf{CTA; BOI} & Corporate Transparency Act, 31 U.S.C. 5336, and the beneficial ownership information reporting regime it created, administered by FinCEN and now limited to foreign-formed reporting companies.\\\hline
\textbf{FCRA; CRA} & Fair Credit Reporting Act, 15 U.S.C. 1681 et seq., governing consumer reports; a consumer reporting agency is a person that assembles or evaluates consumer information to furnish such reports.\\\hline
\textbf{Regulation B; ECOA} & 12 C.F.R. Part 1002, implementing the Equal Credit Opportunity Act, 15 U.S.C. 1691 et seq., including adverse-action notice requirements for business credit applicants.\\\hline
\textbf{Regulation P; GLBA} & 12 C.F.R. Part 1016, implementing the privacy provisions of the Gramm-Leach-Bliley Act, 15 U.S.C. 6801 et seq., governing disclosure of consumers\textquotesingle{} nonpublic personal information.\\\hline
\textbf{Section 1033} & Section 1033 of the Dodd-Frank Act, 12 U.S.C. 5533, under which the CFPB issued the Personal Financial Data Rights rule, 12 C.F.R. Part 1033, on consumer-authorized access to financial account data.\\\hline
\textbf{SDN List; 50 Percent Rule} & OFAC\textquotesingle{}s List of Specially Designated Nationals and Blocked Persons; OFAC\textquotesingle{}s guidance that an entity owned 50 percent or more, directly or indirectly and in the aggregate, by blocked persons is itself blocked whether or not listed.\\\hline
\textbf{SP 800-63} & NIST Special Publication 800-63, Digital Identity Guidelines; Revision 4, including SP 800-63A-4 on identity proofing, was issued July 31, 2025.\\\hline
\textbf{DLDV; DL/ID} & Driver License Data Verification, the AAMVA service that returns match flags comparing submitted driver license or identification card (DL/ID) fields with issuing-jurisdiction records, without releasing the jurisdiction\textquotesingle{}s data.\\\hline
\textbf{TIN; EIN; TIN Matching; Form 1099-K; Form W-9} & Taxpayer identification number; Employer Identification Number, the TIN assigned to entities; the IRS program that confirms whether a name and TIN pair matches IRS records; the information return payment settlement entities file for payees; the form on which a payee certifies its name and TIN.\\\hline
\textbf{Nacha Operating Rules; ACH} & The rules governing the Automated Clearing House network, the U.S. batch electronic funds transfer system.\\\hline
\textbf{C.F.R.; U.S.C.; Fed. Reg.} & Code of Federal Regulations; United States Code; Federal Register, the daily publication of proposed and final rules.\\\hline
\textbf{Industry and document terms} & \textbf{Meaning in this paper}\\\hline
\textbf{ISO} & Independent sales organization, a third party that markets and services merchant accounts on behalf of an acquirer.\\\hline
\textbf{Sponsor bank} & The member bank of a card network under whose membership an acquirer, ISO, or payment facilitator operates, and which remains responsible to the network and its regulators for that activity.\\\hline
\textbf{DBA} & Doing business as: a trade name under which an entity operates that differs from its legal name.\\\hline
\textbf{API} & Application programming interface: a defined software interface through which one system requests actions or data from another.\\\hline
\textbf{AI; agent} & Artificial intelligence; an agent is software that proposes or performs actions on behalf of a principal, which in this architecture acts only under accepted authority.\\\hline
\textbf{mDL; VDC} & Mobile driver\textquotesingle{}s license, a state-issued digital credential; verifiable digital credential, a digitally signed credential whose authenticity can be verified cryptographically.\\\hline
\textbf{Hold; Review; Allow with Controls; final Allow} & Controlled dispositions of policy evaluation. Hold suspends pending evidence or process; Review routes to a permitted decision process; Allow with Controls records a favorable evaluation that cannot proceed until required controls complete; final Allow is the state that permits the consequential action.\\\hline
\textbf{\href{https://chainit.com/product/identity-verification/}{ChainIT ID}; \href{https://chainit.com/product/business-verification/}{ChainIT Org ID}; \href{https://chainit.com/product/digital-pay/}{ChainIT Pay}} & ChainIT\textquotesingle{}s verified individual identity, verified organizational identity, and payment execution capabilities, respectively.\\\hline
\textbf{Compliance Protocol; \href{https://chainit.com/insights/chainit-launches-provable-authority-as-u-s-generative-ai-enabled-fraud-losses-are-projected-to-reach-40-billion-by-2027/}{Authority Protocol}; \href{https://chainit.com/white-papers/}{Transaction Truth} Architecture} & The ChainIT protocol that produces purpose-specific compliance evidence and decisions (CIT-PC-WP-4.1); the separate ChainIT protocol that resolves authority for a session or exact transaction (CIT-PA-WP-7.0); and the umbrella architecture of which both are part.\\\hline
\textbf{Evidence Assertion} & A sourced observation or institutional state represented as a discrete, attributable record (an Evidence Assertion \href{https://chainit.com/product/data-tokenization/}{VDT}), preserving source class, identity, timestamps, limits, and permitted disclosure.\\\hline
\textbf{CIT document identifiers} & ChainIT publication codes: DTT, Distance from the Truth; PC, \href{https://chainit.com/white-papers/}{Provable Compliance}; PA, \href{https://chainit.com/insights/chainit-launches-provable-authority-as-u-s-generative-ai-enabled-fraud-losses-are-projected-to-reach-40-billion-by-2027/}{Provable Authority}; WP, white paper; the trailing number is the version.\\\hline
\textbf{Scientia Potentia Est II, LLC} & The ChainIT affiliate that owns the intellectual property, patent applications, and publications referenced in this paper.\\\hline
\end{longtable}

\clearpage
\unnumbered{References and Source Notes}
\vspace{-0.73bp}{\fontsize{9bp}{12.48bp}\selectfont\color{cgray}External references below are primary-source materials reviewed on 24 September 2026; references [23] through [35], added in version 2.1, were reviewed on 16 and 29 September 2026. Law, guidance, network standards, and source-program conditions have different legal effects. The controlling legal text, contractual edition, jurisdiction, and deployment remain determinative. Reference links point to the publisher; linked public pages may subsequently change.\par}
\vspace{0.7bp}\begingroup\fontsize{9bp}{11.04bp}\selectfont\setlength{\topskip}{8.88bp}

\endgroup
\subsection*{\color{dgreen}ChainIT architectural and technical source basis}
\begingroup\fontsize{9bp}{11.04bp}\selectfont

\endgroup
{\fontsize{9bp}{12.48bp}\selectfont\color{cgray}\textbf{Interpretive boundary.} The illustrative cases, relationship-control inventory, application-level receipt, instruction-version checks, and pilot measurements in this paper are proposed merchant-underwriting applications. They do not represent measured production outcomes, certify compliance, or assert that any patent claim covers the complete combination.\par}


\begin{thebibliography}{99}
\bibitem[C2]{c2} Scientia Potentia Est II, LLC, project identity application materials. Portable Self Sovereign Digital Identity with Immutable Trust \& Privacy-Preserving Verification, particularly paragraphs [0003], [0012]-[0014]. Technical source support for source-based identity, selective disclosure, and portable verification; not deployment verification or a claim-scope opinion.
\end{thebibliography}
\end{document}